\documentclass[%
 reprint,
 amsmath,amssymb,
 aps,
]{revtex4-2}
\usepackage{todonotes}
\usepackage{graphicx}
\usepackage{dcolumn}
\usepackage{bm}
\usepackage{makecell}
\usepackage{comment}

\begin{document}

\preprint{APS/123-QED}

\title{A Neutrino Detector Design for Safeguarding Small Modular Reactors}

\author{Emma Houston}
 \email{ehoust10@vols.utk.edu}
 \author{Sandra Bogetic}
 \affiliation{University of Tennessee, Knoxville.}
 \author{Oluwatomi Akindele}%
\author{Marc Bergevin}
\author{Adam Bernstein}
\author{Steven Dazley}
\affiliation{Lawrence Livermore National Laboratory}

\date{\today}

\begin{abstract}

\end{abstract}
\begin{abstract}
Nuclear reactors have long been a favored source for antineutrino measurements for estimates of power and burnup. With appropriate detector parameters and background rejection, an estimate of the reactor power can be derived from the measured antineutrino event rate. Antineutrino detectors are potentially attractive as a safeguards technology that can monitor reactor operations and thermal power from a distance. Advanced reactors have diverse features that  may present challenges for current safeguards methods. By comparison, neutrino detectors offer complementary features, including a remote, continuous, unattended, and near-real-time monitoring capability, that may make them useful for safeguarding certain classes of advanced reactors. This study investigates the minimum depth and size of an antineutrino detector for a SMR to meet safeguards needs for advanced reactors. Extrapolating performance from several prior reactor antineutrino experiments, this study uses an analytical approach to  develop a possible design for a remote antineutrino-based monitoring device. 

\end{abstract}

\maketitle

\section{Introduction}
The highly penetrating nature of neutrinos and their high rate of production in reactors, makes them attractive as a potential reactor safeguards monitoring technology. Several applications have been identified as potential avenues for further exploration to determine if neutrino detectors can be deployed for safeguard measurements \cite{IAEAsafeguards}. A working group between the International Atomic Energy Agency (IAEA) and antineutrinos experts determined that antineutrino detectors may offer a complementary solution to the current safeguards structure \cite{IAEAsafeguards}. The implementation of neutrino detectors for safeguarding advanced reactors was also a use case finding in the NuTools Study. This assessment was based on expert opinions, including advanced reactor developers, on applications of neutrino detection \cite{Akindele_2021}. Several advanced reactor types, including Small Modular Reactors (SMRs) and Molten Salt Reactors (MSRs) were specifically identified as an opportunity for potential neutrino detector applications. Previous studies on antineutrino detection sensitivities to diversion of thorium-fueled MSRs have been performed, while investigations into SMRs are less abundant \cite{akindele2016antineutrino}. When choosing a detector size, economics and construction feasibility should be considered. With this consideration, the ultimate goal of this study is to identify the smallest and shallowest possible detector to meet IAEA detection goals. 

\section{Background}
\subsection{Neutrino Detectors at Power Reactors}
Nuclear beta decay is the primary driver for neutrino emissions in nuclear reactors. The fission of $^{235}$U and $^{239}$Pu produces neutron-rich fission products that are likely to beta decay. On average, these isotopes undergo $\sim6$ beta decays per fission before reaching stability
\cite{Bernstein_2020}. The neutrino detection rate is proportional to the reactor's thermal output. With a sufficient number of events, the energy spectrum and number of emitted neutrinos can be used to estimate the fissile content of reactor fuel \cite{Bernstein_2020}. However, this study will solely investigate reactor power and operation as a function of neutrino rate, due to the relatively modest number of events that can be acquired for relevant standoff distances and dwell times. 

Several previous experiments have used these neutrino rates from power reactors to determine neutrino properties  \cite{PhysRevD.62.072002, carr2015measurements, https://doi.org/10.48550/arxiv.1003.1391, cao2016overview}.  These neutrino experiments rely on inverse beta decay (IBD) interactions in gadolinium-loaded liquid scintillator to detect neutrinos.

\[ {\bar{\nu}_e} + p \longrightarrow e^+ + n\]

In this paper we will rely on previous gadolinium-loaded liquid scintillator reactor experiments to analytically determine the feasibility of a neutrino power monitoring detector for a generic Light Water SMR, similar to NuScale.

\subsection{Challenges in SMR Safeguards}
Nuclear reactors provide approximately 20\% of the United States electrical output \cite{eia}. Moving to carbon-free sources of electricity requires nuclear energy but the last decade has shown the challenges behind our current generation of nuclear reactors in terms of cost and timeliness of construction. Additionally, the current generation of large scale nuclear plants are well-suited for specific markets but not necessarily adaptable for all facets of the electric grid. Several novel designs have been proposed that differ in fuel type, coolant type, and size. The SMR design has adaptable electrical output, dependent on the number of cores, which can be tailored to the needs of the grid at a particular location \cite{LOCATELLI201475}. This offers the availability to put these reactors in more remote locations. 

SMRs present both advantages and challenges to present day safeguard techniques. On one hand, SMRs generally have smaller cores making them a less attractive target for material theft or diversion. However, the anticipated advantage of remote deployment is a challenge to performing inspections \cite{whitlock2014proliferation}. The flexibility in power generation make SMRs ideal for remote locations or developing infrastructure, but also the number of cores increases the frequency of inspections and the burden on IAEA inspectors\cite{prasad2015nonproliferation}.   

The introduction of neutrino detectors as a remote monitoring tool for SMRs may help address some of these challenges. The IAEA has used remote monitoring technologies for the past few decades for other containment and surveillance (C$\&$S) equipment \cite{osti_10171940}.  As neutrino detectors advance through technology readiness levels with ongoing experiments and are capable of remote operation, neutrino detectors show promise as an additional monitoring tool in the current C$\&$S toolbox \cite{akindele2023exclusion}.

\section{Diversion and Misuse Scenarios}

To design a neutrino detector with the goal of advanced reactor safeguards, it is useful to formulate diversion and misuse scenarios where its use could provide potential indicators of diversion. Two hypothetical diversion and misuse scenarios are defined below. The requirement to reduce total site thermal power is a common element in both scenarios. The relevant neutrino detector signature will be a signal rate decrease due to a reduction in power. 

\subsection{Scenario 1: Recovery from Unexpected Gaps in C$\&$S Measurements}
The increased number of cores at SMR sites, and increased construction of SMRs due to power demands, may increase the burden on the IAEA in-terms of inspector-days needed for inventory and information verification, and/or re-establishing Continuity of Knowledge following an anomaly, These new circumstances may increase the opportunities for the failure of standards safeguards measures. 

Current IAEA safeguards at reactors rely primarily on C$\&$S in the form of optical surveillance or seals. These elements can fail in a variety of ways, including loss of power to surveillance cameras, degradation due to environmental or radiative effects, and other potentially purposeful attacks.  Additionally, some C$\&$S technologies rely on in-person inspection of seals that retrospectively indicate an item has been tampered with, and thus don’t provide real-time monitoring of diversion. Further, many current C$\&$S safeguards methods must be performed for the inspection of fresh fuel in each core at each refueling. This increases the number of inspections required to safeguard a multi-core SMR facility.

In our first concept of operations, the neutrino detector would monitor the aggregate site reactor power, flagging anomalies such as the temporary shutdown of a single core, or power reductions across several cores. As the IAEA considers designs for SMR-specific safeguards concepts, a neutrino-based component offers a complementary capability for continuous online monitoring of aggregate reactor operations, and may help alleviate some of the above problems. Currently, inspections are chosen at random, or based on the other C$\&$S mechanisms in place. The indications and irregularities in the neutrino signal can potentially aid in the decision to pursue unplanned inspections or even routine inspection. Increased deployment has been identified as a challenge behind SMR safeguards \cite{prasad2015nonproliferation}. Remote neutrino-based power monitoring may also help inform decisions about which cores to inspect and prioritize within a larger number of operating reactors. 

\subsection{Scenario 2: Reducing the Reactor Power To Improve Plutonium Quality}
 At the beginning of cycle, the quality of the $^{239}$Pu to other Pu isotopes ratio is higher than at the end of cycle. This is because the beginning of cycle produces $^{239}$Pu through neutron capture on $^{238}$U.  However, as  $^{239}$ Pu is created, it is converted to $^{240}$Pu through neutron capture. Operating at a lower power can slow the production of the higher Pu isotopes such as $^{240}$Pu.  Higher ratios of $^{239}$Pu to other Pu isotopes would be desired by a proliferator because the other Pu isotopes can be challenging to handle to design a device. $^{240}$Pu has a large amount of both decay heat and spontaneous fission neutrons, $^{241}$Pu decays to $^{241}$Am, which is very radioactive. Additionally, $^{242}$Pu has a high rate of spontaneous neutrons and a large critical mass \cite{pellaud2002proliferation}. For these reasons, the ratio of $^{239}$Pu to other isotopes is a key factor in determine the attractiveness of the material to a proliferator.  Most current safeguard approaches for reactors use a combination of C$\&$S and inspections. As a consequence, the production of weapons-grade plutonium through the cycle can remain undetected between inspections. In addition to providing  indications of potential misuse, neutrino detectors may be used to provide redundancy to current C$\&$S and thereby help to deter misuse in these instances.

\begin{figure}
\centering
\includegraphics[width=0.5\textwidth]{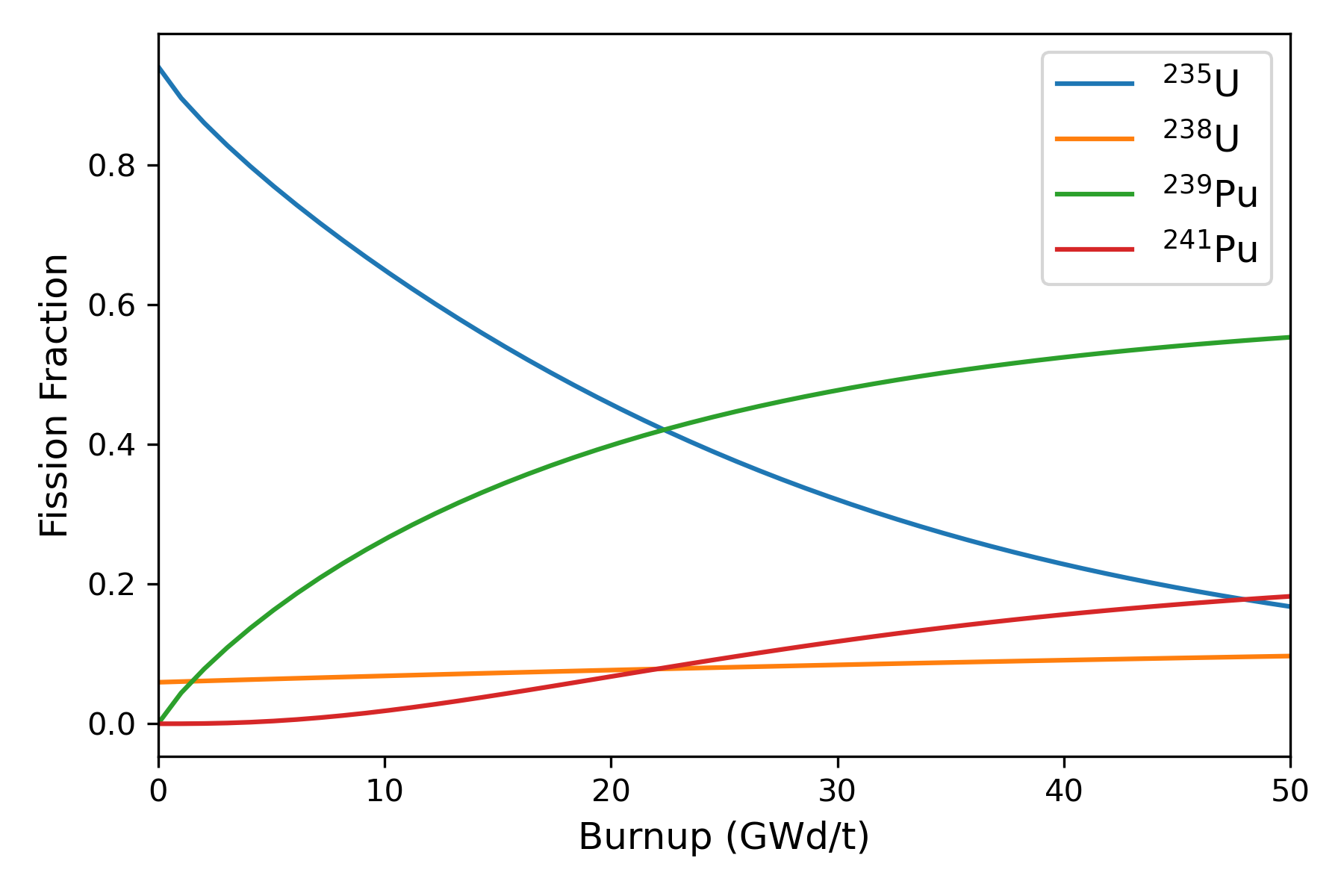}
\caption{\label{fig:burnup} shows the decrease in $^{239}$Pu compared to other fission fractions within the cycle using the approximation for fission fractions shown in \cite{MILLS20202130} for a 3$wt\%$ LEU fueled PWR. The decrease in $^{239}$Pu to total Pu ratio is caused by neutron capture and the conversion of $^{239}$Pu to $^{240}$Pu. While this graph does not include the relative amounts of $^{240}$Pu in the cycle, the ratio of $^{239}$Pu decreases compared to the other Pu isotopes. Further comparisons of the simplified fission fraction approximations to reactor data are shown in the following \cite{bogetic2023anti}. }
\end{figure}

Neutrino detectors provide a method to monitor reactor power levels from a standoff distance outside of the site perimeter.  Neutrinos are impossible to shield and thus adversaries cannot hide the signals arising from reactor power operation. These neutrino measurements may provide valuable information about the period over which power has been decreased, which may give insight into the quality of plutonium that could be produced from a reactor. 

In both scenarios, the goal of the neutrino detector is to detect the reduction in a timely manner. This reduction in neutrino signal is indicative of the reduction in power of the aggregate site power.

\section{Signal and Background Determination}

\subsection{Reactor Signal}
The detector mass and its distance from the reactor core affect the signal rate. Here we vary these detector parameters to identify a decrease in reactor power consistent with a single reactor shutdown from a six core generic SMR configuration.  Considering the goal of this study is to analyze a detector’s safeguard capabilities from outside of the site perimeter, detector parameters to meet current IAEA safeguards standards are investigated.  The Exclusion Area Boundary (EAB) and Low Population Zone (LPZ) are found using dose estimates, among other factors, to require limiting distances from the reactor to the EABs and LPZs \cite{regguide1195}.  There is significant cost reduction for SMRs due to their smaller EABs \cite{osti_1168629}. The smaller EABs in turn makes SMRs attractive sites for neutrino detectors, which can be much closer to the reactor and remain outside the perimeter compared to traditional Light Water Reactors (LWRs). We define the distance from the core to the EAB as the minimum distance of the detector from the reactor to be considered non-intrusive. 

The antineutrino signal can be found using Eq. \eqref{eq:1}. It is assumed the reactor power is constant thus the number of fissions over time does not change. A more robust approximation would include calculations with regard to burnup over time. However, this study is intended to reflect the averaged detection rates over a whole cycle. 

\begin{equation} \label{eq:1}
N_{det}(E_{\nu}, t) = \frac{\varepsilon(E_{\nu})}{4\pi L^2}\phi_{\nu}(E_{\nu}, t) \sigma(E_\nu)N_T P_{surv}(E_\nu, L)
\end{equation}

In Eq. \ref{eq:1}, $N_{det}$ is the number of neutrinos detected per second \cite{bernstein2020colloquium}. The estimated detection efficiency, $\varepsilon$ is $80 \%$, based on recent experiments \cite{carr2015measurements}, \cite{An_2012}, and was treated as the average over all neutrino energies. The estimated distance from the reactor to the detector is L. The flux averaged cross section, $\sigma$, is found by \cite{abe2012reactor}.  This study assumed an operating thermal power of 160 MWt per core \cite{nu-scaledesign} with 6 cores operating during normal operation. The oscillation dependent survival probability, $P_{surv}$, is given by Eq. 65.9 in the Review of Particle Physics \cite{PhysRevD.98.030001}. The energy-dependent neutrino flux is given a fifth order polynomial with coefficients for $^{235}$U, $^{238}$U, $^{239}$Pu, and $^{241}$Pu in the following references \cite{mueller2011improved}, \cite{huber2016reactor}. The fission fractions and energy per fission from the following source was used \cite{li2022scalability}. 
$N_T$ is the number of free protons within the detector and is directly proportional to the detector mass. Similar to previous antineutrino detector experiments, we will assume the scintillator will be Gd-loaded linear alkyl benzene (LAB) and thus has a free proton per molecule of 30 and a density of 0.86 g/ml \cite{yeh2007gadolinium}. This is used to calculate the number of free protons per ton.

\subsection{Background Interpolation}
This study considers three background types: $^{9}$Li-$^{8}$He production, fast-neutron production, and accidentals.

\subsubsection{Muon-Induced Backgrounds}
 Both cosmogenic radionuclide production ($^{9}$Li and $^{8}$He), and fast-neutron production are induced by atmospheric muons. When muons interact with the $^{12}$C inside the scintillator, $^9$Li and $^8$He can be produced. Subsequent $\beta-n$ decays of these isotopes mimic antineutrino-induced IBD interactions \cite{DayaBay, maesano2012lithium}. Fast neutrons can be produced when muons interact with nuclei in the rock surrounding the detector. Some of these fast neutrons then produce recoil protons in the detection medium, creating a prompt signal, then thermalize within the detector and capture to create a delayed signal. The combination of these two correlated signals and their temporal structure again mimics the IBD interaction\cite{maesano2012lithium}. To reduce these backgrounds, the detector will require some overburden to reduce the muon flux.
 However, as with detector size, detector depth has a strong impact on detector construction and operational costs.

To approximate muon-induced backgrounds, several reactor neutrino experiments were benchmarked and an analytical model was used to estimate background rates. Table\ref{tab:table1} displays the data used to approximate the background from $^{9}$Li, $^{8}$He and fast-neutrons.

\begin{table*}
\caption{\label{tab:table1} shows Li-He, fast-neutron, and accidental backgrounds from previous neutrino detector experiments. These values are used to estimate the production of muogenic and accidental backgrounds. 
}  
\begin{ruledtabular}
\begin{tabular}{lcccc}
\textrm{Experiment}&
\textrm{Depth mwe}&
\textrm{Li-He (events/day)}&
\textrm{Fast Neutrons (events/day)}&
\textrm{Accidentals (events/day)}\\
\colrule

    
     RENO & 120 & $12.45 \pm 5.93$ \cite{ahn2012observation} & $5.0 \pm 0.13$ \cite{ahn2012observation} & $4.3 \pm 0.06$  \cite{PhysRevLett.108.191802}\\

     Double Chooz & 120 & $5.01 \pm 1.43 $\cite{schoppmann2017latest}& $3.42 \pm 0.23$ \cite{schoppmann2017latest} & $0.344 \pm 0.002$ \cite{schoppmann2017latest}\\
     
     Daya Bay & 250 & $3.1 \pm 1.6$ \cite{ahn2012observation} & $0.84 \pm 0.28$ \cite{ahn2012observation} & $9.85 \pm 0.06$ \cite{ahn2012observation}\\

     Daya Bay & 265 & $1.8 \pm 1.1$ \cite{ahn2012observation} & $0.74 \pm 0.44$ \cite{ahn2012observation} & 
     $7.67 \pm 0.05$  \cite{ahn2012observation}\\

     Double Chooz & 300 & $0.97_{-0.16}^{+0.41}$ \cite{schoppmann2017latest} & $0.586 \pm 0.061$ \cite{schoppmann2017latest} & $0.106 \pm 0.002$  \cite{schoppmann2017latest}\\

     RENO & 450 & $2.59 \pm 0.75$ \cite{ahn2012observation} &$0.97 \pm 0.06$ \cite{ahn2012observation} & $0.68 \pm 0.03$  \cite{PhysRevLett.108.191802}\\

     Daya Bay & 860 & $0.16 \pm 0.11$ \cite{ahn2012observation} & $0.04 \pm 0.04$ \cite{ahn2012observation} & 
     $3.25\pm 0.03$  \cite{ahn2012observation}\\
\end{tabular}
\end{ruledtabular}
\end{table*}


The background rates in the theoretical detector are then assumed to scale with the volume of the detector in the case of $^{9}$Li, $^{8}$He and surface area in the case of fast neutron backgrounds. The muon track lengths in the detector are proportional to the detector volume,increasing the number of number of potential $^{12}$C atoms to interact with. The rates for fast neutrons are scaled to the surface area of the detector as the surrounding rock scales directly with this type of background. 
The detector dimensions used to determine the surface area and volume are shown in Table \ref{tab:table3}. Daya Bay has multiple detectors at the same depth, each with their own background measurements. In these cases the average background was used, rather than treating all detectors as individual data points. For the Li-He background fit, the Daya Bay backgrounds for 250 mwe and 265  mwe  were averaged to avoid over fitting. (This was not needed for fast-neutron background fitting as the fast neutrons backgrounds at these two depths were very similar.) 
Lastly, RENO has more recent lower backgrounds measurements that are the result of more stringent multiplicity measurements \cite{RENOrecent}. However, the larger backgrounds measurements were chosen in this approximation for conservatism. 

\begin{table*}
\caption{\label{tab:table3}%
The detector and reactor characteristics of each experiment are listed by increasing depth. The radi, height, baselines, and signal are averaged amongst detectors from the same experiment in the same hall. 
}
\begin{ruledtabular}
\begin{tabular}{lcccccccc}
\textrm{Experiment}&
\textrm{Depth}&
\textrm{Mass}&
\textrm{Radius}&
\textrm{Height}&
\textrm{Site Power} &
\textrm{Baseline} &
\textrm{Efficiency}&
\textrm{Signal}\\

\textrm{} &
\textrm{ mwe }&
\textrm{tons}&
\textrm{cm}&
\textrm{cm}&
\textrm{GWt}&
\textrm{m} &
\textrm{$\%$}&
\textrm{Events/Day}\\

\colrule
     
     RENO & 120  & 16 \cite{ahn2012observation} & 137.5 \cite{park2013production} & 315.0 \cite{park2013production} & 16.52 \cite{ahn2012observation} & 408.56 \cite{ahn2012observation} & $64.70 \%$ \cite{PhysRevLett.108.191802} & $779.05$ \cite{PhysRevLett.108.191802}\\

     Double Chooz & 120  & 9 \cite{queval2010characterization} & 115.0 \cite{ardellier2006double}& 245.8 \cite{ardellier2006double} & 8.5 \cite{schoppmann2017latest} & 400 \cite{crespo2015double} & - & 293.4 \cite{schoppmann2017latest}\\

     Daya Bay & 250 & 20 \cite{An_2012} & 155 \cite{an2016detector} & 308.2 \cite{an2016detector}& 17.4 \cite{an2017measurement} & 512 \cite{an2014search} &  $78.8\%$ \cite{ahn2012observation} &716.0 \cite{ahn2012observation}\\

     Daya Bay & 265  & 20 \cite{An_2012} & 155 \cite{an2016detector} & 308.4 \cite{an2016detector}& 17.4 \cite{an2017measurement} & 561 \cite{an2014search} &$78.8\%$ \cite{ahn2012observation} &532.29 \cite{ahn2012observation}\\

     Double Chooz & 300 & 9 \cite{queval2010characterization}& 115.0 \cite{ardellier2006double} & 245.8 \cite{ardellier2006double} & 8.5 \cite{schoppmann2017latest} & 1050 \cite{crespo2015double} & $78.0 \%$ \cite{abe2014improved}\footnote{Double Chooz's selection efficiency and Gd-capture fraction are listed separately in the following reference \cite{abe2014improved}. It is assumed these can be treated as follows to obtain the total efficiency: (91.5$\% \times 85.3 \% = 78.0 \%$).} & $39.0$ \cite{schoppmann2017latest} \\

     RENO & 450  & 16 \cite{ahn2012observation}& 137.5 \cite{park2013production} & 315.0 \cite{park2013production} & 16.52 \cite{ahn2012observation} & 1443.99 \cite{ahn2012observation} & $74.50\%$ \cite{PhysRevLett.108.191802} & $72.78$ \cite{PhysRevLett.108.191802} \\

     Daya Bay & 860 & 20 \cite{An_2012} & 155 \cite{an2016detector} & 308.5 \cite{an2016detector} & 17.4 \cite{an2017measurement} & 1579 \cite{an2014search} & $78.8\%$ \cite{ahn2012observation} & 70.7 \cite{ahn2012observation}\\
     \end{tabular}
\end{ruledtabular}
\end{table*}
Daya Bay, Double Chooz, and RENO muon-induced background data were each fit to an exponential function shown by Eq. \ref{eq:2}. Each of these detectors have unique selection criteria to optimize the signal to background rates. These exponential fits were averaged for both types of correlated backgrounds. 

\begin{equation} \label{eq:2}
y = me^{-tx} 
\end{equation}

\begin{figure}[]
\centering
\includegraphics[width=0.5\textwidth]{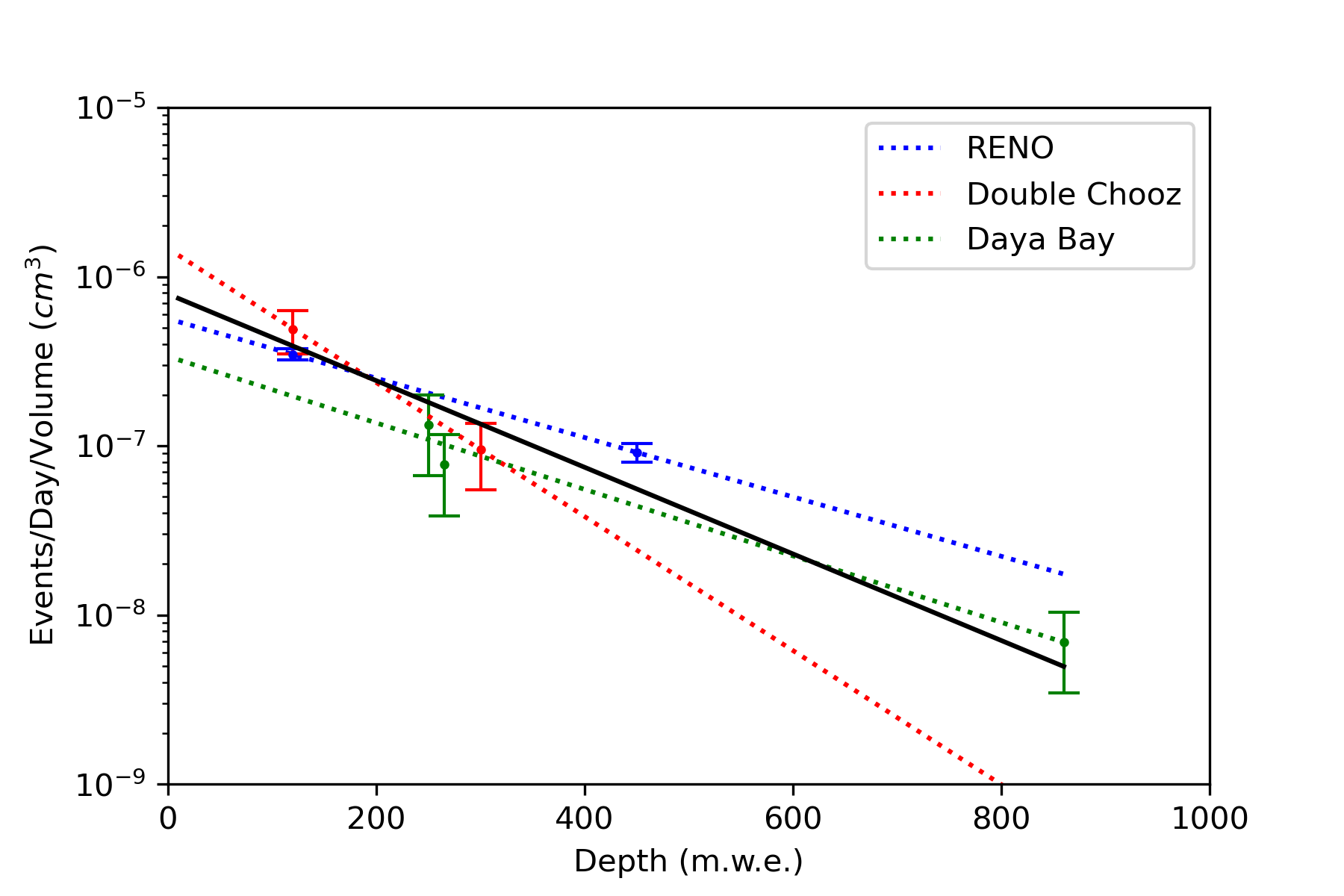}
\caption{\label{fig:Li-He}The equations of best fit for these background measurements are: $y = 7.89 \times 10^{-7}e^{-5.90 \times 10^{-3}x} \frac{events}{day-cm^3}$ for the Li-He induced background, where x is depth in  mwe . Backgrounds for each set up detectors are fit to Eq. \ref{eq:2} and then averaged. Fig. \ref{fig:Li-He} show this functions that are then sampled from to approximate background.}
\end{figure}

\begin{figure}[]
\centering
\includegraphics[width=0.5\textwidth]{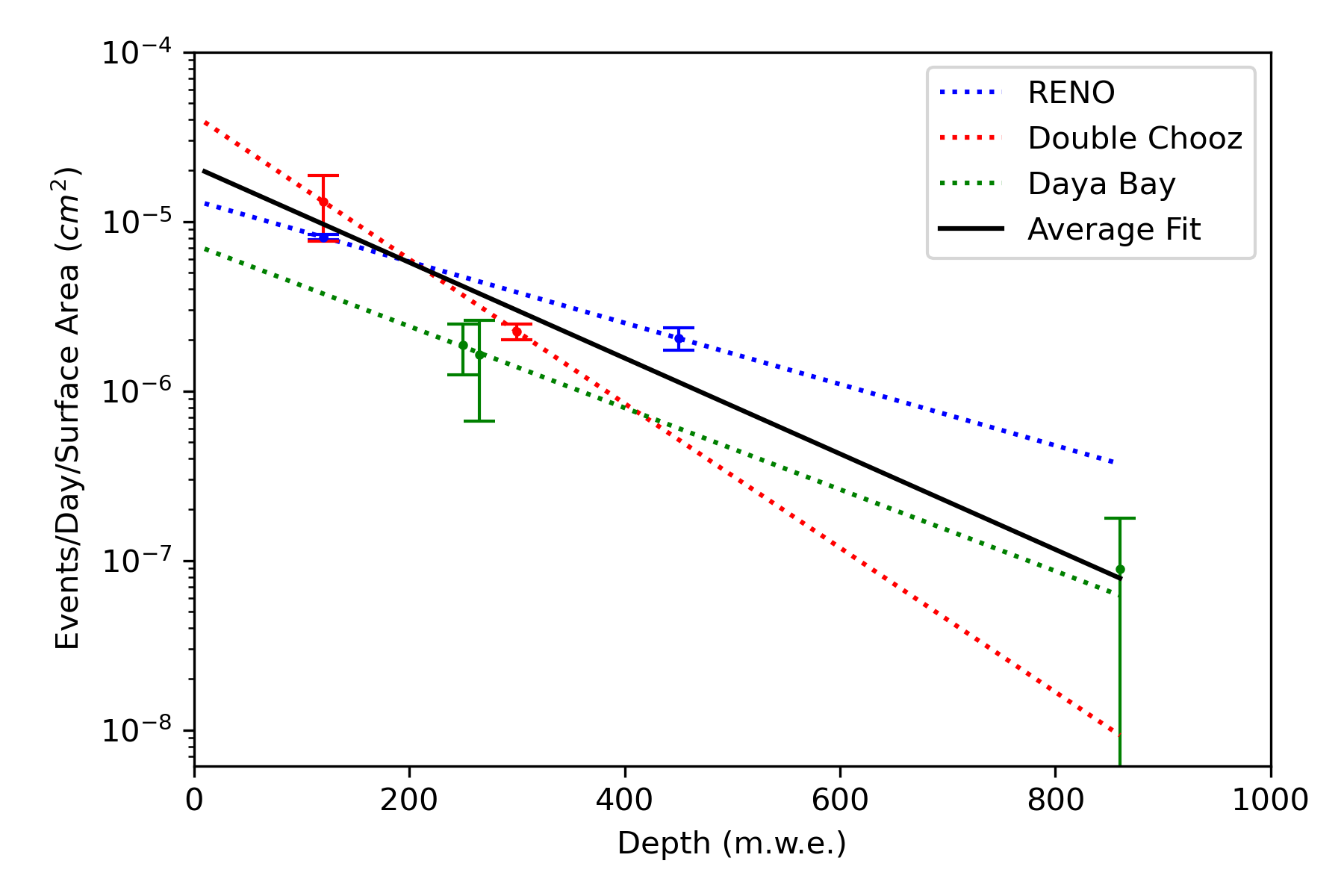}
\caption{\label{fig:fastn}The equations of best fit for these background measurements are $y = 2.10 \times 10^{-5}e^{-6.50 \times 10^{-3}x} \frac{events}{day-cm^2}$ for the fast-neutron induced background, where x is depth in mwe . Backgrounds for each set up detectors are fit to Eq. \ref{eq:2} and then averaged. Fig. \ref{fig:fastn} show this function that are then sampled from to approximate background.}
\end{figure}


There are several considerations when extrapolating this data to approximate background rates at shallower depths. This method for background approximations does not account for veto capabilities of the varying detectors.  Additionally, this method assumes that these backgrounds are directly related to the attenuated muon flux through the rock overburden, and thus Eq. \ref{eq:2} can be applied. Higher energy muons are found at deeper rock overburdens because lower energy muons are more likely to interact and stop entirely. At shallow depths, both the muon energy and angular distribution have a significant impact on muon intensity however this study approximates muon flux using solely depth \cite{bogdanova2006cosmic}. 

Lastly, RENO reported a contamination of $^{252}$Cf in their detector, with a more significant impact on the far RENO detector than near. A multiplicity requirement helps to mitigate the effects of the $^{252}$Cf  source, causing the contamination of the near detector to be negligible and the contamination of the far detector to amount to $0.095 \pm 0.018$ events/day \cite{RENOCfSource}. Shallower depth detector experiments, such as Palo Verde at 32 mwe, may offer some insight into muon-induced backgrounds at shallower depths \cite{PhysRevD.62.072002, piepke2002final}. However, given the significant differences in detector designs, solely monolithic detector experiments were chosen for extrapolation. 

\subsubsection{Accidentals}
Aside from the muon-induced backgrounds, accidentals, or uncorrelated backgrounds, are another important source of IBD-like events in a detector. This background is not depth dependent, and can be caused by a variety of factors including detector contamination, selection criteria, shield thickness, and detector size. For these reasons, accidentals can not be correlated to a single parameter, such as depth in the previous case, and must be treated separately.


The causes for uncertainty in accidentals can be the uncertainties radioimpurities of the detector materials or surrounding rock overburden, or any other natural radioactivity that can be attributed to coincidence. Ways to minimize these accidental backgrounds include shielding thickness, stringent multiplicity measurements in space, time, and energy, and general detector cleanliness. Due to the unpredictability of accidental backgrounds, a scaled prediction was not created and rather the largest accidentals rate from the detector measurements was used. Like fast neutrons, accidentals are also proportional to the surface area surrounding a detector. The largest accidentals rate per surface area from the detectors using Tables \ref{tab:table1} and \ref{tab:table3} is from the far RENO detector at 450 mwe. The maximum number of accidentals divided by surface area is $5.57 \times 10^{-5} \frac{events}{day-cm^2}$. As this analysis will further show, a surface area is found for each of the detector masses and multiplied by this value to find a conservative estimate for accidental backgrounds in our theoretical detector. 

\subsection{Uncertainty Estimates}
The Double Chooz detector signal absolute uncertainty, including uncertainties in cross section, fractional fission rates, thermal power, mean energy per fission, and distance, is propagated to be 1.7  \% \cite{carr2015measurements}. Neglecting the effect of burnup has the consequence of degrading the precision of the thermal power estimate. Therefore the thermal power component of the uncertainty was increased from 0.5\% to 5\%, while keeping all of the other uncertainties constant. 
The updated signal uncertainty then propagated to 5.25\%.

In all events, the largest systematic uncertainty percentage from each respective data set, cosmogenic radionuclides ($^{9}$Li and $^{8}$He), fast-neutrons and accidentals, were used when generating assumed systematic uncertainties from background. The largest cosmogenic radionuclide systematic uncertainty percentage was 50\% from Daya Bay at 250 mwe and the largest fast-neutron uncertainty percentage was 100 \% from Daya Bay at 860 mwe shown in Table \ref{tab:table1}. The largest uncertainty percentage was $4.8 \%$ from Palo Verde. These maximum uncertainties using previous experiment data in Table \ref{tab:table1} will be used as a conservative approximation of background uncertainties.

Using these systematic uncertainties from previous detector experiments, 4 trials were conducted using combinations of $\pm$ reactor signal systematic uncertainty and $\pm$ background systematic uncertainty. Using the combinations of Gaussian distributed total signal rates with these systematically shifted means, there was no significant different in detection times, false alarm rates, and missed alarm rates. Given this, consideration of systematic uncertainties was not included in the final detection times and alarm rates.

\section{Detection Metrics}
\subsection{IAEA Requirements for Containment and Surveillance}
With established signal and background rates that are detector depth, size, and distance dependent, we can determine how quickly we can detect a reduction in power as discussed in the diversion scenarios. The time to detect the reduction in power from 6 to 5 operating reactors, and the Type I and II error rates \cite{international2022iaea} are the metrics considered for the success of the detector. The Type I error is the frequency of which the null hypothesis is declared to be false when it is true. Type II error is the frequency of which the null hypothesis is declared true when it is false \cite{international2022iaea}. The Type I and II error rates will be further described as the false and missed alarm rates, respectively. The IAEA timeliness detection goals depend on the quality and quantity of fissile material \cite{international2022iaea}. One objective is to reduce the time required to detect the diversion of a significant quantity of fissile material.  
Although the IAEA timeliness component of the IAEA inspections goals may vary by reactor \cite{international2022iaea}, the desired goal to observe a reduction in power was determined to be 30 days in this study for conservatism. This should not be used as a sole indicator of diversion, and thus 30 days would allow ample time for further inspections. These goals govern the technical requirements for detector depth and size.  The false alarm probability, $\alpha$, is the frequency that there is an indication that diversion has occurred when there has been no diversion. The goal for IAEA false alarms (Type I error) is generally small \cite{international2022iaea}, and therefore $\alpha <= 0.05$ is the goal used in this study.  The Type II error goal, or missed alarm rate, was set to be 10 $\%$  Therefore the targets for this study are $\alpha = 0.05$ and $\beta = 0.1$.

\subsection{Simulating Signal}
\begin{figure}[!htbp]
\centering
\includegraphics[width=0.5\textwidth]{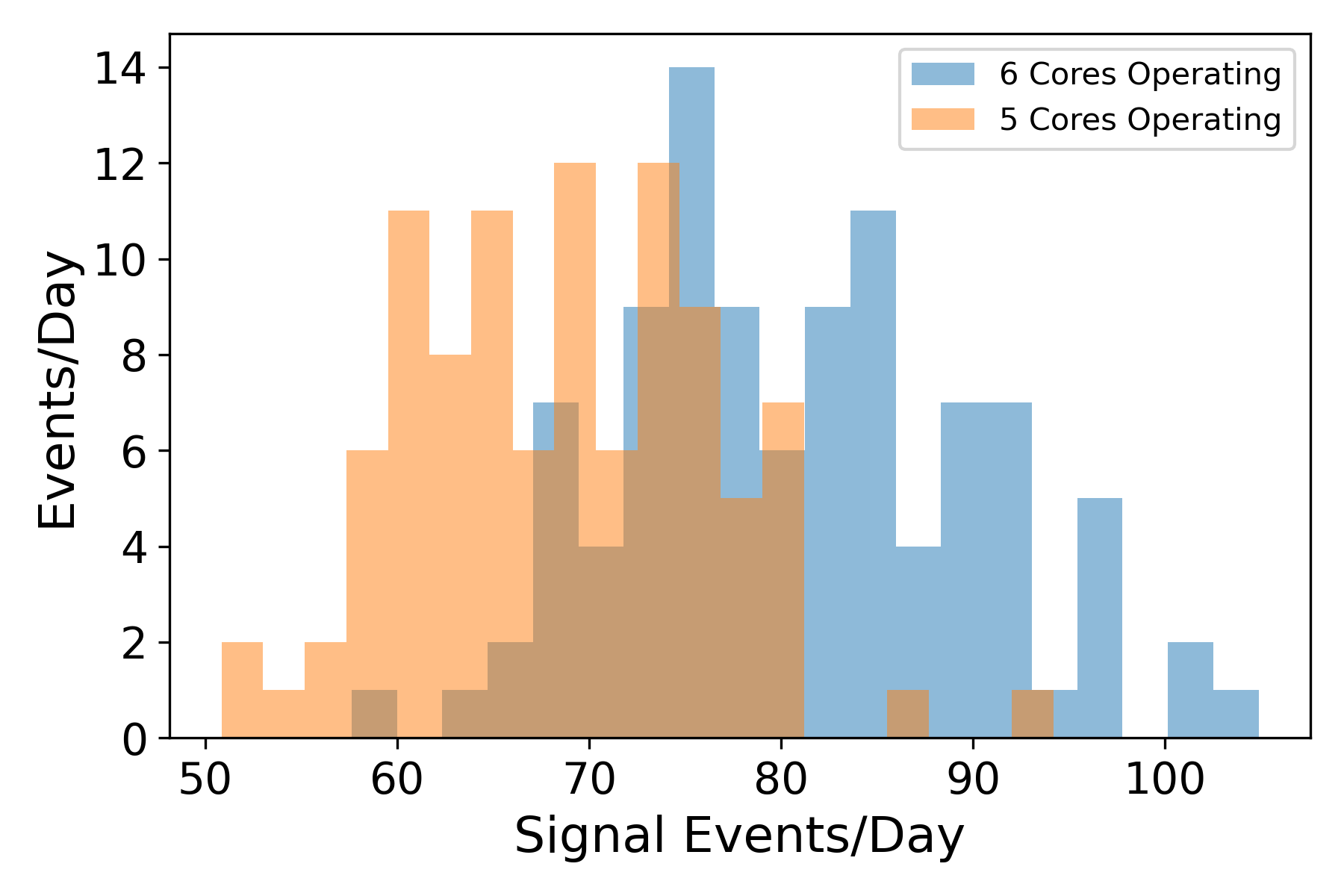}
\caption{\label{fig:guass}  The randomly Guassian-like distribution of signal rates for 100 days steady state reactor power and a reduced power, in a 5 ton detector at a depth of 20 mwe. The decreased power is representative of either of the diversion scenarios occurring.}
\end{figure}

To identify the detection time for the reduction in power of a reactor, the total signal from two scenarios is simulated. In the first, the expected detector response is normally distributed around the average signal from 6 operating cores and the corresponding background at a given depth for 200 days. In the second instance, the detector response is normally distributed around the signal from 6 operating cores and corresponding background for 100 days and then the detector response is simulated for 5 operating reactors and the same background for the following 100 days. To create this data, the Numpy random sampling feature was utilized \cite{harris2020array}. Fig. \ref{fig:guass} shows the Guassian-like distribution of the number of signal events per day distributed data using the Numpy package. This is for 6 reactors operating at 160 MWth each at a distance of 244 m from the reactors. The following data and analysis is for a 5 ton detector at 20 mwe, however this was repeated for various configurations. 

When this data is sampled, especially in instances with smaller detectors, there is a reasonable possibility that the signal randomly sampled from the normal distribution will be negative. In these cases, the code re-samples from the normal distribution until a realistically feasible value is chosen. However, this is frequently not necessary as the vast majority of sampled signals are positive values. 

\begin{figure*}[!htbp]
\centering
\includegraphics[width=0.9\textwidth]{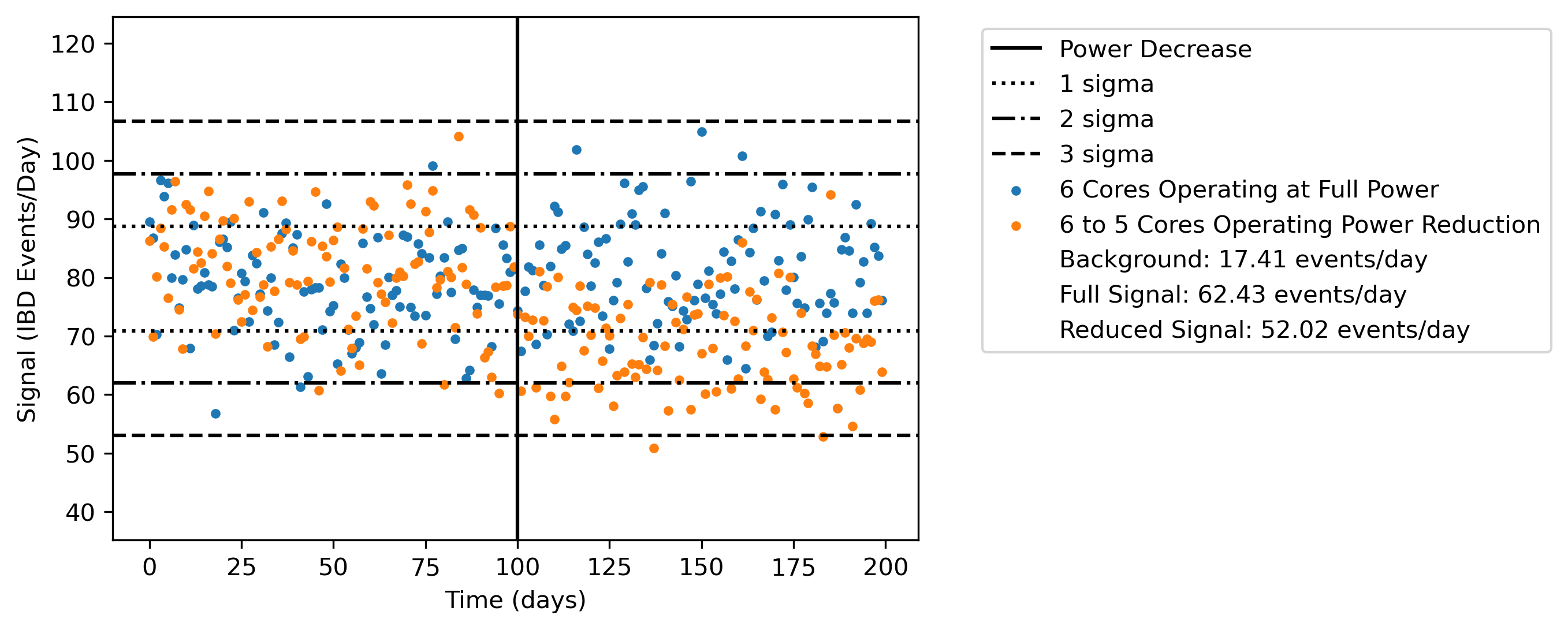}
\caption{\label{fig:sampled} shows the Gaussian distributed signal over time for the diversion and non-diversion case. The vertical line indicates when there is a power decrease, and marks the time at which diversion has occurred. 
}
\end{figure*}

Fig. \ref{fig:sampled} shows both the diversion and non-diversion simulated signal. Using solely the neutrino signals on a per day basis, it can be difficult to differentiate between signals in the non-diversion and diversion simulated data. Alarming techniques are employed to be able to find how quickly after diversion detection can occur. To lessen the inherent fluctuations in the data, a rolling sum of the data is found for a specified "window".  This window can be optimized for each detector configurations, but in this case, a time window of 10 days was used. The rolling sum of this data is then subtracted from the expected sum (the full power signal expectation multiplied by the window). This expected sum is found using a calibration window of 20 days, where a full power signal is Gaussian-distributed and sampled. The mean of this calibration period is then used as $s_{exp}$. This calibration period is used to emulate the time before a detector is put into operation where the expected signal is found. Eq. \eqref{plot_eq} shows this method for data analysis where the time window is defined as $x$ in days, the expected full power signal is $s_{exp}$, the actual signal is $s_{a}$, and $t$ is time in days.  

\begin{equation} \label{plot_eq}
Diff(t, x) = x \times s_{exp} - \sum_{i = t-x}^t s_{a} (t) 
\end{equation}
It is expected that in the first few weeks of operation, the expected signal will be determined and used as a comparison to the remainder of operation. An alarm indicates when the difference in this aggregate signal crosses 3 sigma above the expected difference of aggregate signals.

 If this alarm is greater than 30 days, the time to detection goal in this study, then this is considered a missed alarm. If this occurs prior to the reduction in power, then this is considered a false alarm. There can be multiple false alarms for each trial as there can be multiple data points above the 3-sigma threshold. There cannot be multiple missed alarms per trial as the data can only miss the detection goal once. This is repeated for $10^5$ trials and the time to detect the reduction in reactor power is averaged, and the frequency of missed and false alarms are found. False alarms are found on a per day, per trial basis whereas missed alarms are solely found per trial. False alarms can be remediated by requiring sequential data to be above the 3 sigma threshold, however this should optimized dependent on the detector configuration. 

\begin{figure*}[!htbp]
\centering
\includegraphics[width=0.9\textwidth]{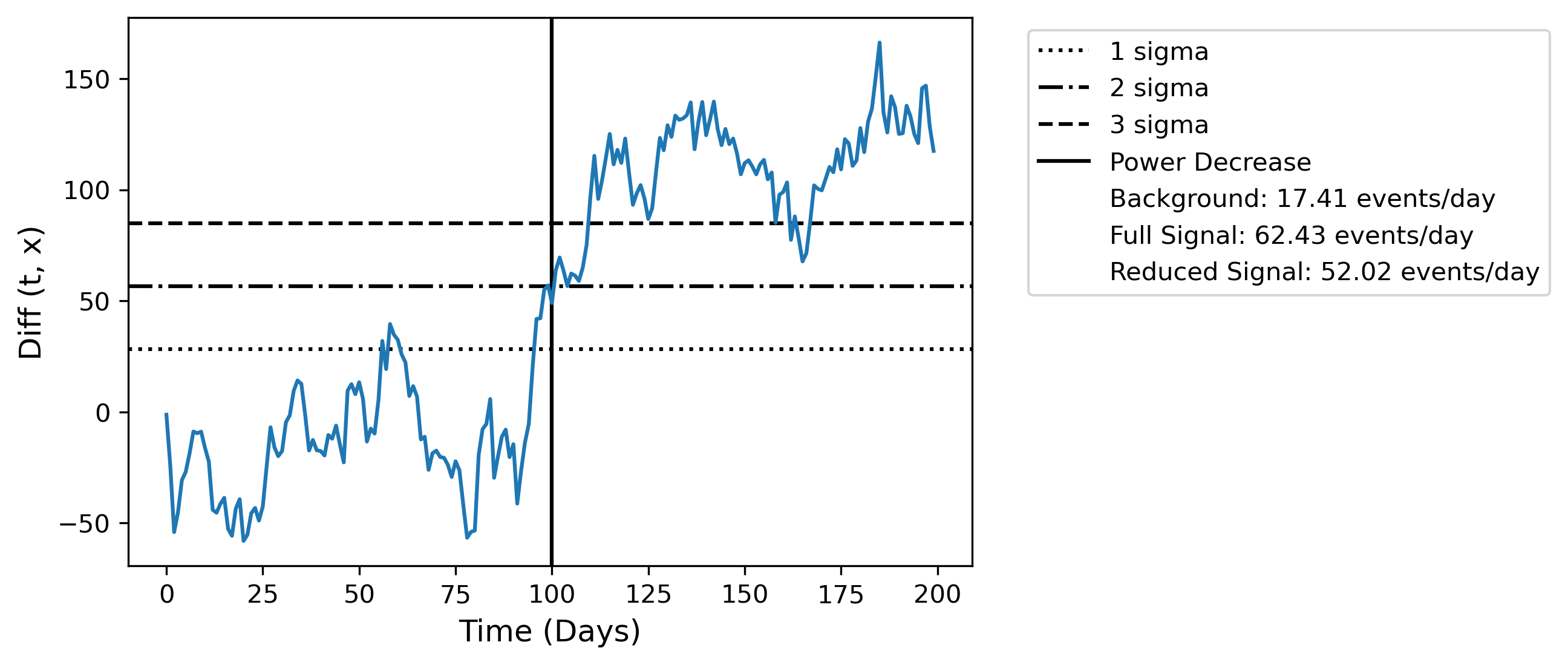}
\caption{\label{fig:agg}shows the $Diff(t,x)$ as defined by Eq. \eqref{plot_eq} versus time in days. The first 10 days, the specified time window in this case, are voided as there is not sufficient previous simulated data to find the aggregate counts from the previous days. When this difference in signal crosses the threshold for 3 sigma, an alarm is recorded.}
\end{figure*}


 In this example detector shown in Fig. \ref{fig:sampled}, at 5 tons at a depth of 20 mwe, the detection time is $\sim$ 10 days, $\alpha = 0.006$ and $\beta = 0.024$. These values are well within the IAEA guidance for the detection of diversion. Of course, this process would show an indication that diversion may have occurred and should not be used as the sole tool for diversion. However, this should encourage further inspections of a facility or be used in the prioritization process of facility inspection. Additionally, this mechanism is complementary to traditional material accountancy procedures to ensure fuel is not immediately replaced following diversion. 

This process was then repeated for detectors with fiducial volumes of 1 to 7 tons, and at depths of 20 mwe to 140 mwe in increments of 20. Table \ref{tab:table5} shows the the time to detect the reduction in power beyond the case described in detail above. Table \ref{tab:table6} shows the false alarm rates for these detectors and Table \ref{tab:table7} shows the missed alarm rates. The sample standard deviation shown in Eq. \ref{standev} is used to find the standard deviation of the time to detection, where $x_i$ is each time within a trial, $\bar{x}$ is the mean time of the trial, and n is the number of trials ($10^5$).

\begin{equation} \label{standev}
    \sigma = \sqrt{ \frac{\sum_{i=1}^n (x_i - \bar{x})^2}{n -1}}
\end{equation}

\begin{table*}[htb!]
\caption{\label{tab:table5} shows, in days,  the average time over $10^5$ trials to detect a reduction in power from 6 to 5 operating cores 3 sigma confidence and the standard deviation found by Eq. \ref{standev} .} 
\begin{ruledtabular}
\begin{tabular}{cccccccc}
\textrm{Depth}&
\textrm{1 ton}&
\textrm{2 tons}&
\textrm{3 tons}&
\textrm{4 tons}&
\textrm{5 tons}&
\textrm{6 tons}&
\textrm{7 tons}\\
\hline     
20 mwe & 33.1 $\pm$ 24.9	&23.4 $\pm$ 20.3&	16.6 $\pm$ 15.1&	12.4 $\pm$ 10.7&	10.0 $\pm$ 7.5	&8.5 $\pm$ 5.3& 7.6 $\pm$ 4.0 \\ 
40 mwe & 33.0 $\pm$ 25.0	&23.1 $\pm$ 20.2&	16.4 $\pm$ 15.0&	12.3 $\pm$ 10.6&	9.8 $\pm$ 7.2&	8.4 $\pm$ 5.2&	7.5 $\pm$ 3.9 \\ 
60 mwe & 33.0 $\pm$ 25.0	&23.0 $\pm$ 20.0&	16.3 $\pm$ 14.8&	12.2 $\pm$ 10.4&	9.8 $\pm$ 7.1&	8.4 $\pm$ 5.1&	7.5 $\pm$ 3.9 \\ 
80 mwe & 32.8 $\pm$ 25.0	&22.9 $\pm$ 20.0&	16.1 $\pm$ 14.6&	12.1 $\pm$ 10.1&	9.7 $\pm$ 7.1&	8.3 $\pm$ 5.0&	7.4 $\pm$ 3.7 \\ 
100 mwe & 32.8 $\pm$ 24.9	&22.9 $\pm$ 20.0&	16.0 $\pm$ 14.4&	12.0 $\pm$ 10.1&	9.7 $\pm$ 7.0& 8.3 $\pm$ 4.9&	7.4 $\pm$ 3.7 \\ 
120 mwe & 32.6 $\pm$ 24.9	&22.6 $\pm$ 19.7&	15.9 $\pm$ 14.5&	11.9 $\pm$ 10.1&	9.6 $\pm$ 6.9&	8.2 $\pm$ 4.9&	7.4 $\pm$ 3.7 \\ 
140 mwe & 32.6 $\pm$ 24.9	&22.5 $\pm$ 19.7&	15.8 $\pm$ 14.3&	11.8 $\pm$ 9.9&	9.6 $\pm$ 6.8&	8.2 $\pm$ 4.8&	7.3 $\pm$ 3.7 \\

\end{tabular}
\end{ruledtabular}
\end{table*}

\begin{table*}[htb!]
\caption{\label{tab:table6}
The fractional false alarm rates show how frequently the aggregate signal crosses the alarm threshold without a simulated reduction in power. All of the following fractional false alarm rates are within the IAEA goals for C$\&$S metrics and thus not a factor when determining the minimum size and depth detector. }
\begin{ruledtabular}
\begin{tabular}{cccccccc}
 Depth &1 ton&2 tons&3 tons
 &4 tons &5 tons&6 tons&7 tons\\
\hline
20 mwe &0.005 &	0.006 &0.006 &	0.006 &	0.006&	0.006 &	0.006  \\ 
40 mwe& 0.005 &	0.005 &0.006 &  0.006 &0.006 &	0.006 &	0.006  \\ 
60 mwe& 0.005 &	0.005 &0.006 &	0.006 &0.006 &	0.006 &	0.006  \\ 
80 mwe &0.005 &	0.005 &0.006 &	0.006 &0.006 &	0.006 &	0.006  \\ 
100 mwe &0.005 &0.006 &	0.006 &	0.006 &0.006 &  0.006 & 0.006  \\ 
120 mwe & 0.005 &0.006 &0.006 &	0.006 &0.006 &	0.006 &	0.006  \\ 
140 mwe & 0.005 &0.006&	0.006 &	0.006 &0.006 &	0.006 &	0.006  \\ 
\end{tabular}
\end{ruledtabular}
\end{table*}

\begin{table*}[htb!]
\caption{\label{tab:table7}
The missed alarm rates are indicative of the fractional rate that the detection times are after the goal detection time (30 days), in which case the reduction in power was not detected soon enough.  }
\begin{ruledtabular}
\begin{tabular}{cccccccc}
 Depth &1 ton&2 tons&3 tons
 &4 tons &5 tons&6 tons&7 tons\\
\hline
20 mwe & 0.554	&0.299&	0.142&	0.06&	0.024&	0.009&	0.003 \\ 
40 mwe & 0.551&	0.296	&0.138	&0.059	&0.023&	0.008&	0.003 \\ 
60 mwe & 0.548&	0.291	&0.136	&0.057	&0.022&	0.008&	0.003 \\ 
80 mwe & 0.546&	0.288	&0.133	&0.054	&0.021&	0.008&	0.003 \\ 
100 mwe & 0.543&0.286	&0.128	&0.055	&0.02	&0.007&	0.002 \\ 
120 mwe & 0.541	&0.281	&0.128	&0.053	&0.02	&0.007&	0.002 \\ 
140 mwe & 0.54	&0.278	&0.125&	0.052&	0.019&	0.006&	0.002 \\ 
\end{tabular}
\end{ruledtabular}
\end{table*}

\section{Conclusion}
This exercise is entirely reactor power dependent to make determinations about the steady state and reduced power signal. The diversifying advanced reactor market requires new methods to safeguard nuclear material. As an example, the current safeguards regime for LWRs is not designed for advanced fuel types like molten salt reactors with liquid fuel and pebble bed reactors \cite{peel2022nuclear}. Neutrino detectors could be used as a way to determine the operational status of these reactors, where traditional item accountancy may pose challenges.  This method can be applied to a range of advanced reactors. 
As discussed previously, this paper determines the background and signal rate using analytical methods and data from previous neutrino experiments to provide a benchmark. 

Depth likely has a minimal impact on the detection time and metrics as the remaining 5 cores dominate the remaining signal rather than depth-dependent muon-induced backgrounds. 

The two diversion and misuse scenarios discussed in this paper are the intentional shutdown of one core of six to divert fuel or the decrease in power of all six cores to improve the ratio of Pu-239 in used fuel. For the latter, it is assumed that all cores are reduced to 83 \% of full operating power making both of these scenarios appear identical with respect to neutrino signals. Of course, reactors could operate at an even lower capacity to slow other Pu isotopes production for longer. This would lead to even faster detection times and better results with respect to the IAEA requirements.

Neutrino detectors can assist the IAEA with some of the anticipated challenges behind SMR safeguards including increased deployment and remoteness of reactors. The World Nuclear Association reports that around 30 countries are beginning to develop nuclear programs \cite{wn}. This will inevitably create increased costs and burden for the IAEA to safeguard these new reactors. Neutrino detectors located outside of the reactor sites may offer the capability for the IAEA to monitor these reactors remotely and offer an additional layer of redundancy to the current C$\&$S safeguard regime. 

Simulations using RAT(-PAC) can be used to compare these monte-carlo driven predictions for signal and background to the analytical methods displayed in this work \cite{askinssnowmass2021}. 
Neutrino detectors are capable of measuring reactor power from distances outside of a site boundary. This research investigated varying detector sizes and depths to compare to IAEA metrics for success for C$\&$S. Our results show that a 5 ton detector with an overburden of 20 mwe was able to detect the reduction in power from 6 operating reactors to 5 in $\sim$ 10 days with a false and missed alarm rate of  $\alpha = 0.006$ and $\beta = 0.024$, respectively. Other methods of alarming systems may allow for smaller and shallower detectors. This would reduce the cost of construction and operation. 

\section*{Acknowledgement}
This material is based upon work supported by the Department of Energy National Nuclear Security Administration under Award Number DE-NA0000979 through the Nuclear Science and Security Consortium. 
Lawrence Livermore National Laboratory is operated by Lawrence Livermore National Security, LLC, for the U.S. Department of Energy, National Nuclear Security Administration under Contract DE-AC52-07NA27344. LLNL-JRNL-852145.

\bibliographystyle{ieeetr}
\bibliography{Ref}

\end{document}